\documentstyle[12pt]{article} \textheight=24.5cm \textwidth=16cm \topmargin=-1.5cm \begin{document} \begin{center}
\large{ON THE VORTEX-POINT CHARGE COMPOSITE:  \\
CLASSICAL ORBITS AND QUANTUM BOUND STATES}  
\end{center}
\vskip 2cm
Subir Ghosh\\
Physics Department, Gobardanga Hindu College,\\
24 Pgs. (N), West Bengal, India.\\
\vskip 3cm

\noindent Abstract\\
The possibility of composite systems arising out of a point charge
interacting with a Nielsen-Olesen vortex in 2+1-dimensions is 
investigated. It is shown that classical bounded orbits are possible
for certain ranges of parameters. Long lived metastable states are
shown to exist, in a semi-classical approach, from the study of the
effective potential. Loss of self-adjointness of the Hamiltonian and
its subsequent self-adjoint extension in some cases leads to bound
states.

\newpage
\section{Introduction}

2+1-Dimensional physics has come of age in the last couple of decades. Specifically
with the advent of anyons \cite{wil} , (excitations having arbitrary spin and
statistics), and its posited presence ranging from diverse condensed matter
systems such as high $T_c$ superconductivity, fractional quantum Hall effect to
more exotic scenarios in the high energy regime such as processes in the presence
of cosmic strings etc.,
2+1-dimensions has ceased to be just a laboratory for testing ideas conceived
for the "physical" 3+1-dimensions. The present paper deals with studying the
energy spectrum of a quantum (charged) particle in the presence of a 
Nielsen-Olesen magnetic vortex
\cite{no} , in  2+1-dimensions. 

The motivation is twofold. The physical existence of magnetic vortex lines
in type II superconductors \cite{sch} (in an external magnetic field) makes
the study of a vortex-particle system interesting. Also according to the present
lore and as envisaged by Wilczek \cite{wil}, a possible realisation of anyons
in nature is some sort of a composite object, consisting of a magnetic flux
tube (of the "fictitious" Chern Simons gauge field) attached to a bose or a
fermi particle. Albeit the crucial difference in nature of the two magnetic
fields, the existence of (quantum) bound states, (as we demonstrate here),
should lend credibility to the hypothetical anyon structure.

Let us elaborate a little on our system, our modes of analysis as well as
the results that we have obtained. The system is that of a non-relativistic
point charge in the presence of a vortex, the latter being the Nielsen-
Olesen vortex solution in the Abelian Higgs Model (AHM) \cite{no, raj}. We
restrict ourselves to minimal gauge invariant coupling between the point
charge and the gauge field.

We start with a thorough discussion on the classical aspect of the
problem, which turns out to be quite tricky. We show that {\it classical
bounded trajectories} \cite{gold} {\it of the particle are allowed for certain ranges of the
parameters of the model}. This analysis is important since it
hints at the possible existence of quantum bound states, which brings us
to the Schrodinger equation problem of the charge in presence of a
classical vortex potential. One would be too optimistic to conclude that
upon quantization, the closed orbits would correspond to bound states.
Apparantly this is not the case. Essentially this is because the 
Hamiltonian in question being {\it positive definite}, {\it negative
energy bound states can not appear}. Still we will show that the effective
potential energy profile is such that it allows {\it metastable states of
quite long life time}. Previous works \cite{ss} in a simlar vein are those
of a point particle in the presence of a 't Hooft-Polyakov monopole in
3+1-dimensions.

However, {\it bound states in the strict sense do exist}, for certain values
of effective angular momentum at least. This is related to the {\it loss
of self adjointness of the Hamiltonian, due to the singular vortex
potential, and its subsequent self adjoint extension}, \cite{rs, 
cal}, which allows only the above bound states. An early work in this
connection is \cite{case}.

The paper is organised as follows: in section 2, we introduce the vortex potential
and the detailed classical dynamics. In section 3, we discuss the metastable
quantum bound states. Section 4 deals with the problem of self adjointness
and the ensuing true bound states. We conclude the paper in section 5. 

\section{Classical dynamics}

The Nielsen-Olesen vortex solutions \cite{no} have provided a deep connection
between a relativistic field theory, (the AHM), and the system of type II
superconductors. If the superconducting material is invariant along the $z$-direction,
the system is essentially a two-dimensional one, in the $x-y$-plane \cite{raj}.
In the former case vortices appear in the two space dimensional slice of the
magnetic flux lines extended along the third direction. Hence one can study
the vortex solutions in the 2+1-dimensional AHM.

Let us start by introducing the vortex solutions of the AHM \cite{no, raj}.
The ($c=1$) Lagrangian is
\begin{equation} 
L_{AHM}=-{1\over 4}F^{\mu\nu}F_{\mu\nu} + {1\over 2}(D_\mu \phi)^*
(D^\mu \phi) - {\lambda\over 4}(\mid\phi\mid ^2 -F^2)^2, \label{eq1} 
\end{equation}
where $F^{\mu\nu}=\partial^\mu A^\nu -\partial^\nu A^\mu$, $D^\mu \phi=\partial^\mu
\phi-ieA^\mu\phi$. The problem has a cylindrical symmetry and the asymptotic
solution, (i.e., the vortex), in polar $(r,\theta)$ coordinates is
\begin{equation} 
A_r=A_0=0,~~ A_\theta (r)\approx{n\over{er}}+{{\alpha e^{-eFr}}
\over{\sqrt r}}, ~ r\to\infty, \label{eq2}
\end{equation}
\begin{equation}
\phi=e^{in\theta}(F+\beta e^{-\sqrt{\lambda}Fr}).\label{eq3}
\end{equation}
We will be confined to the $n=1$ or single vortex line or 'flux-tube' sector,
since numerical work has indicated it to be stable. $\alpha$ and $\beta$ are
constants and $\alpha$ is of the order of ${\sqrt{eF}\over {e}}$.

In this work we only consider the particle to be interacting with the gauge
potential $A^\mu$. Obviously, other interactions such as Yukawa or vector 
couplings with $\phi$ can be studied as well. Henceforth we consider the
gauge field to be external. Clearly this particular solution of $A^\mu$
in (\ref{eq2}) generates no electric field. The magnetic field is
\begin{equation}
B={1\over e}2\pi\delta^{(2)}(\vec r)+\alpha({1\over{2r}}-eF){e^{-eFr}\over{\sqrt{r}}}
\label{eq4}
\end{equation}
The matter Lagrangian with $U(1)$ gauge invariant coupling is
\begin{equation}
L_{matter}={1\over 2}M\dot{r}^2+{1\over2}Mr^2{\dot\theta}^2 +j^\mu A_\mu,\label{eq5}
\end{equation}
where $j^\mu$ is the conserved particle current.
We denote time and space derivatives of $O$ by $\dot O$ and $O'$ respectively.
 Writing explicitly the
potential, we obtain

\begin{equation}
L_{matter}={1\over 2}M\dot {r}^2+{1\over 2}Mr^2{\dot\theta}^2 +gr\dot\theta({1\over{er}}+
{\alpha e^{-eFr}\over{\sqrt r}})
,\label{eq6}
\end{equation}
Note that the term $gr\dot\theta({1\over{er}})={g\over e}\dot\theta$ is a total
time derivative and does not influence the classical equations of motion.
It is dropped in the present classical analysis.
The Lorentz equation for the particle is
\begin{equation}
M\ddot{X}^i=g\epsilon^{ij}\dot {X}^j B, \label{eq7}
\end{equation}
where $M$ and $g$ are mass and charge of the particle and $B$ is without the
$\delta$-function term. 
In polar coordinates we
have
\begin{equation}
\dot{\vec X}=\dot {r}\hat {e}_r +r\dot\theta \hat {e}_\theta, \label{eq8}
\end{equation}
\begin{equation}
\ddot{\vec X}=(\ddot r-r{\dot\theta}^2)\hat {e}_r +(r\ddot\theta +2\dot r\dot\theta )\hat {e}_\theta ,\label{eq9}
\end{equation}
and the Lorentz equation in component form is
\begin{equation}
M(\ddot r-r{\dot \theta}^2)=gr\dot\theta B,~~ M(r\ddot\theta+2\dot r\dot \theta)=-g\dot rB. \label{eq10}
\end{equation}
Note that ${g\over e}$ has the dimension of $h$. In the quantised version
the particle charge $g$ becomes related to $e$ by $g=k(eh)$ where $k$ is
some number.
The $\theta$ equation following from (\ref{eq6})  is given by
\begin{equation}
Mr^2\dot\theta +\alpha g {\sqrt r}e^{-eFr}=L, \label{eq11}
\end{equation}
where $L$ is a constant. Thus a generalised angular momentum is conserved.
Borrowing from the classical terminology, this is one of the integrals of
motion, the other being the conserved energy, since a static background
interaction is chosen. The $r$-equation provides the energy, when we
substitute $\dot\theta$ from (\ref{eq11})
\begin{equation}
M\ddot r=-{d\over{dr}}[{1\over{2M}}({{L-\alpha g\sqrt{r}e^{-eFr}}\over r})^2],
\label{eq12}
\end{equation}
and the conserved energy $E$ is
\begin{equation}
E={1\over 2}M{\dot r}^2 +{1\over{2M}}
({{L-\alpha g\sqrt{r}e^{-eFr}}\over r})^2=
{1\over 2}M{\dot r}^2 +{1\over{2M}}V(r). \label{eq13}
\end{equation}
This is just the sum of kinetic and effective potential energy.

To proceed further, we must ascertain first that the potenial energy function
$V(r)$ in (\ref{eq13})  does have a profile which is able to sustain bounded classical
motion of the particle. Clearly $V(r)$ is positive infinity at $r=0$ and goes to
zero at large $r$. We want to argue that between $r=0$ and $r=\infty $, there
is one minimum and one maximum, which will give the well.
 If we express the potential energy
$V(r)$ as $V(r)=[T(r)]^2$, then
\begin{equation}
V(r)'=2T(r)T(r)',~~ V(r)''=2(T')^2 +2TT''. \label{eq14}
\end{equation}
Clearly $V'=0$ has two roots, $T=0$ and $T'=0$. It is easy to see that the
root $T=0$ constitutes the minimum. To show that under some restrictions,
the other root $T'=0$ does represent a maximum, $V''=2TT''$ has to be
negative. We evaluate $V''$ explicitly and use $T'=0$ to express it in 
the following form
\begin{equation}
V''\mid_{T'=0}=-{{2(\alpha g)^2e^{-2eFr}}\over {r^2}}({1\over 2}-eFr)[{1\over 2}
-({1\over 2}-eFr)^2]. \label{eq15}
\end{equation}
Hence the allowed range of $r$ is such that either ${1\over 2}>eFr>0$ or
$eFr>{5\over 4}$.

Thus a potential well is formed with the minimum ar $r_1$ obeying
\begin{equation}
L=\alpha g \sqrt{r_1}e^{-eFr_1}, \label{eq16}
\end{equation}
and the maximum at $r_2$ obeying
\begin{equation}
L=\alpha g\sqrt{r_2}e^{-eFr_2}({1\over 2}+eFr_2), \label{eq17}
\end{equation}
provided $r_2>r_1$ and $r_1$ is within the allowed range. Obviously the
minimum value of $V(r)$ is zero, the expression being positive definite.

Let us now turn to numerical results. Since the equations for the minimum
and maximum, (\ref{eq16}), (\ref{eq17}), involve transcendental functions, we have given graphical
solutions in Fig. 1 with the potential well strucure in Fig. 2, for the 
particular solution chosen in Fig. 1. From Fig. 1 one can see that there is a
lower bound for the gradiant of the straight line (II), below which there is
no solution for $r_1$. However, for this extreme value of the parameter,
solution for $r_2$ is such that $r_2<r_1$. On the other hand, there are solutions
for $r_1$ and $r_2$, with $r_2>r_1$, which is required for well formation,
 such that $r_1$ is at least less than
${1\over {eF}}$, that is the relevant dimensionless variable $eFr_1<1$. In Fig. 1,
(II) and (III) give the values $eFr_1=0.08$, $eFr_2=0.21$. In Fig. 2, we have
plotted the dimensionless variables $V(r)/{(LeF)^2\over{2M}}$ vs. $eFr$. The
well is formed within the range of the vortex, where the exponential damping
can compete with the 
 ${L^2\over{2Mr^2}}$ centrifugal term. But in this
region the asymptotic solution may not be too reliable.

We now give a rough idea of how the system might look close to the origin. One
can have solutions of $A^\mu$ and $\phi$ field equations, (obtained from (1)), 
such that
$$
A_0=A_r=0,~~ A_\theta =\alpha eF^2 r+\beta e^3 F^4 r^3,~~ \phi=CeF^2 r +pe^3 F^4 r^3,
$$
with $\alpha=-{\Lambda\over 6},~ \beta ={\Lambda^2\over {24(1+4\Lambda)}},
~C^2=-{\Lambda^2\over{3(1+4\Lambda)}},~ p=-{{\Lambda C}\over 12},~ \Lambda =
{\lambda\over e^2}.$
The field equations are satisfied to $O((eFr)^n)$, $n>4$. Comparing with
\cite{no}, we find that near the origin, $\phi$ goes to zero correctly. The
magnetic field is 
\begin{equation}
B={{2\lambda F}\over e} +O(r^2).\label{eq18}
\end{equation}
$B$ will behave properly near origin if the $O(r^2)$ term opposes the constant
value, which is true for ${\lambda\over{e^2}}>{1\over 4}$.

Now, the $L_{matter}$ and the energy are
\begin{equation}
L_{matter}={1\over 2}M{\dot r}^2 +{1\over 2}Mr^2{\dot\theta}^2 +gr{\dot\theta}
(\alpha eF^2 r +O(r^3)), \label{eq19}
\end{equation}
\begin{equation}
E={1\over 2}M{\dot r}^2 +{1\over {2Mr^2}}(L-\alpha geF^2 r^2)^2. \label{eq20}
\end{equation} 
Apart from a constant shift of ${{L\alpha geF^2}\over M}$ this potential energy
function is just that of a two dimensional harmonic oscillator, with the minimum
value of zero at $r_1=\sqrt{L/(\alpha geF^2)}$. Hence we note that an 
approximate solution of the gauge field, valid at short distance, provides 
in the leading order, a two 
dimensional circular oscillator potential, which
indeed has stable classical orbits. These two limiting results,
that is gauge fields at large and small $r$, do indicate 
that the system in question is classically stable.

\section{Quantum dynamics: metastable states}

Let us notice at the outset that the Hamiltonian considered by us is of the form

\begin{equation}
H={1\over {2M}}[\vec {p}-g\vec{A}(r)]^2,\label{21}
\end{equation}
where in our case, $\vec {A}(r)$ is the vortex potential \cite{no}. However,
without going into details, for a purely magnetic field,
$H$ is a non-negative, hermitian operator. 
 We consider the Hilbert space to be composed of wave functions which are
square integrable on the plane and regular at the origin. This means that
the energy spectrum is positive and purely contineous, contrary to the nature
of bound states.
We will discuss later
in more detail the Schrodinger equation analysis which also points to the same
conclusion as above. For the present, arguing from a physical point of view,
a potential well which is everywhere positive and goes to zero asymptotically
can not produce a bound state, (for special cases contrary to this conclusion,
see Simon in \cite{rs}),
simply because the particle can tunnel out and live outside the well.
This example clearly shows that
closed classical trajectories, which are present here,
 are in no way a sufficiency condition for
quantum bound states. After the above compelling arguments against the existence
of bound states in the strict sense, (that is states having negative energy), let
us look at the potential profile in Fig. 2 more carefully. We immedietly notice that 
although the well is quite sharp, (since the terms responsible are exponential),
the subsequent fall off towards zero is very gradual, (thanks to the inverse
power law). Obviously this will reduce the tunnelling amplitude to a large extent
and so the states having positive energies that are well inside the "well" can
not live outside the well due to the greater potential energy. Thus the well
quite efficiently traps the particle states and these are termed by us as
metastable states or resonances. Even though the square integrable eigen
functions are disallowed, in the spirit of semiclassical quantization, we
generate harmonic oscillator states about the potential minimum. This will
give the spectrum of metastable states.

It is now straightforward to perform the oscillator approximation near $r_1$.
While considering the asymptotic solutions,
we write $r=r_1 +x$ in $V(r)$ of (\ref{eq13})  and keep upto $O(x^2)$ terms and get
\begin{equation}
V(r_1+x)={1\over{2M(r_1 +x)^2}}[L-\alpha g\sqrt{(r_1 +x)}e^{-eF(r_1 +x)}]^2,
\label{eq22}
\end{equation}
\begin{equation}
\approx {1\over {2Mr_1}^2}[\alpha g\sqrt{r_1}e^{-eFr_1}({1\over {2r_1}}-eF)^2]
x^2. \label{eq23}
\end{equation}
Identifying this energy as ${1\over 2}kx^2$ we find
\begin{equation}
w=\sqrt{{k\over M}}={1\over M}{{\alpha ge^{-eFr_1}}\over \sqrt {r_1}}
\mid {{1\over
{2r_1}}-eF} \mid ={{gB(r_1)}\over M}. \label{eq24}
\end{equation}
Note that this frequency corresponds precisely to the Larmor frequency for
a charged particle in a magnetic field.

Hence the metastable energy spectrum consists of the levels
\begin{equation}
E_n=(n+{1\over 2})\hbar w. \label{eq25}
\end{equation}
(The anharmonic corrections can be included as well).

Near the minimum, the
energy of the system in oscillator approximation becomes
\begin{equation}
E={1\over 2}M\dot{x}^2 +{1\over 2}kx^2 
={1\over {2M}}{p_x}^2 +{1\over 2}kx^2. 
\label{eq26}
\end{equation}
Alternatively the Bohr-Sommerfield condition can be used to obtain the spectrum of
(\ref{eq26}), which is an equation of ellipse in phase space. Thus $\int p_x dx$ is the
area of the ellipse, which is a multiple of $h$. One can also compute the 
phase integral $\int _{a}^{b} p_r dr$ where $a$ and $b$ are the classical turning points
obtained by solving (\ref{eq20})  when $M\dot {r}=0$. However near the potential minimum,
\begin{equation}
\int_a^b p_r dr=\int_a^b \sqrt{2ME} dr=\sqrt{2ME},~(b-a)\approx x.\label{eq27}
\end{equation}
All these will give rise to a spectrum similar as that of (\ref{eq25}).

In case of the solutions valid near the origin, once again expanding near $r_1$
and writing $V(r_1 +x)={1\over 2}x^2$, we find
\begin{equation}
w=\sqrt{k\over M}={gB\over M}\label{eq28}
\end{equation}
with $B$ computed in (\ref{eq18}).
This is also of the same form as (\ref{eq24}). Here we can give the exact energy
levels or wave functions for the point particle.

Indeed, one can do a detailed Bohr-Sommerfield quantisation, by introducing
radial and angular quantum numbers for the $r$ and $\theta$ phase space
integrals. The standard way is to replace ${d\over dt}$ by ${d\over d\theta}$
(using the $\theta$ equation of (\ref{eq11})) in the $r$ equation 
obtained from (\ref{eq6}), in order to
obtain an equation for the orbit \cite{gold}. The energy $E$ appears as a
parameter. Using the orbit equation and the definition of the quantum
numbers, the energy is expressed in terms of the quantum numbers. However, due
to the complicated nature of the equations of motion, we have adopted a more
pedestrian approach to obtain the metastable state energy spectrum in a
harmonic oscillator approximation about the potential minimum.

\section{Quantum dynamics:true bound states}

Let us start this section with some preliminaries, which are distinct for
2+1-dimensions from the conventional 3+1-dimensions, only in some technical
details. For a generic central potential $V(r)$, the complete set of
eigenfunctions are of the form $\psi(r,\theta)=f(r)exp(il\theta )$, with $l$
an integer for singlevaluedness, where $f(r)$ satisfied the radial equation
\begin{equation}
 [{{d^2} \over {dr^2}} +{1\over r}{d\over {dr}} -{l^2\over {r^2}} +{{2M}\over
{\hbar^2}}(E-V(r)]f(r)=0, \label{eq29}
\end{equation}
with $f(0)=0$ \cite{mes}. In our particular case, with the explicit form of
$V(r)$ from (\ref{eq13}),
 we have to solve the radial equation 
\begin{equation}
[{d^2 \over {dr^2}} +{1\over r}{d\over {dr}} -(({L\over r}+{{\alpha g}\over{e
\sqrt r}}e^{-eFr})^2-k^2)]f_{L}(r)  
=0, \label{eq30}
\end{equation}
where $L=l-{g\over e}$, $l\epsilon Z $ and $k^2=2ME\geq 0 $ with $f_{L}(0)=0 $. Note that
here the total time dervative term in (\ref{eq6}) has not been dropped, so
that the full expression of $B$ in (\ref{eq4}) is being considered.

One should not confuse the addition of ${g\over e}$ in the angular momentum
 (which came from the ${g\over e}\dot\theta $ term mentioned above 
in the Lagrangian), with
the ageold recipe of changing particle statistics in the quantum version. In
external field problem angular momentum is quantised in integers even though
the expression differs from the canonical one \cite{jac}. The effect of total 
derivative terms in quantum mechanics is briefly discussed in \cite{wil}.

In our case, with positive $eF$, for large $r$, (that is away from the origin),
the exponential terms die out quickly leaving only the inverse power law
falloff, which controls the large $r$ behaviour.
Now, if the Hilbert space is that of regular square integrable wave functions
on the plane, then the spectrum is purely continuous with $k\geq 0$. In fact the
eigenfunctions asymptotically behave as
\begin{equation}
\psi_L (kr)\approx C_l J_L (kr), \label{eq31}
\end{equation}
where $C_l$ is a constant \cite{mes}. The above analysis 
rigorously shows that bound states in 
general are not possible.

However, one must be more careful in imposing boundary conditions \cite{alb}
on the wave functions due to the presence of the $\delta$-function term in
the magnetic field. This points towards a contact interaction which 
visciates the self adjointness of the Hamiltonian \cite {rs} at the origin. {\it The remedy
is to look for self adjoint extensions (if present) of the Hamiltonian and
this can lead to bound states} \cite{cal} for some restricted set of parameters.

Generally we tend to overlook the difference between Hermitian and self
adjoint operators in quantum physics, but there is a difference in the
structure of their respective domains, which crucially governs the dynamics
in many cases of physical interest. For unbounded operators, (that do not
have bounded expectation values, such as energy, momentum, angular
momentum and position), one has to specify the domain of the operator as well
as its acton on the domain. Without geting involved in too many technical
details, an operator $T$ is called Hermitian if $T\subset {T^*}$, 
that is $D(T)\subset {D(T^*)}$
and $T\phi =T^*\phi $ for all $\phi\epsilon D(T)$. Here the adjoint $T^*$ is
defined as $(T\psi ,\phi )=(\psi, T^*\phi )$ for all $\psi ,\phi\epsilon D(T)$
with $D(T)$ denoting the domain of $T$. $T$ is self-adjoint if $T=T^*$ that is
{\it iff} $T$ is Hermitian and $D(T)=D(T^*)$. {\it Only the self-adjoint operators
can be exponentiated to give one parameter unitary groups which governs the
dynamics in quantum mechanics}. 

The basic criterion of self-adjointness is the following: Suppose $T$ is a
self-adjoint operator and a $\phi $ exists such that $T^*\phi=i\phi $. Then
also $T\phi =i\phi $ and
$$ -i(\phi,\phi )=(i\phi,\phi )=(T\phi,\phi )=(\phi,T^*\phi )=(\phi,T\phi )
=i(\phi,\phi )$$
which implies $\phi =0$ and this is true for $T^*\phi =-i\phi $ as well. Hence for
self-adjointness, it must be ensured that $ker(T^*\pm i)=0$ or $D(T^*)=D(T)$.
However, if $T$ is not self-adjoint on a domain, but has $n_{\pm}$
independent solutions $T^*\phi_i =\pm i\phi_i$ for some $\phi_i\epsilon D(T^*)$
then, only if $n_+ =n_- =n$, one is allowed to make an extension of $T$ to
$\tau$, where $D(T)\subset {D(\tau)}$
and $\tau\phi =T\phi $ for all $\phi\epsilon D(T)$),
such that $\tau$ is self-adjoint. Basically one modifies the domain of $T$ to
$\tau$ such that $D(\tau)=D(\tau^*)$ and this also ensures that $ker(\tau\pm i)=0$. The
vectors $\phi_i$ generate the deficiency subspace and $n_{\pm}$ are referred
as the deficiency indices. Note that for simple differential operators, as
in the present case, $T^*$ is the same as $T$, but acting on $D(T^*)$, which
in general is different from $D(T)$. 

Our task is now to express the Hamiltonian in a suitable form $T$ and 
find independent, normalisable solutions of
\begin{equation}
T\phi_{\pm}^i =\pm i\phi_{\pm}^i, \label{eq32}
\end{equation}

Since we are looking at regions close to the origin, the exponential term
is dropped, compared to more singular centrifugal term.
 Replacing $f(r)$ and $kr$ by ${{u(r)}\over {\sqrt r}}$ and $\rho$ in (\ref{eq29}),
we obtain
\begin{equation}
{{d^2 u_p}\over{d\rho ^2}} +(1-{{p(p+1)}\over {\rho ^2}})u_p =0, \label{eq33}
\end{equation}
where $p(p+1)=-({1\over 4}-L^2)$.
In general the solutions of the Schrodinger equation as well as its first
order partial derivatives will be continuous, uniform and bounded functions
over all space, including $\rho=0$. Here we also have $f(0)=0$. This induces
the following natural boundary conditions on $u_p$ as well;
\begin{equation}
u_p(0)={{du_p(0)}\over {d\rho}}=0. \label{eq34}
\end{equation}
Let us see somewhat heuristically why the singularity problem at $r=0$ affects
only some of the $p$ states. The general solution of (\ref{eq33}) near the origin is
\begin{equation}
u_p\approx A\rho^{p+1} +B\rho^{-p}. \label{35}
\end{equation}
For $p$ different from 0 or -1, to maintain regularity at $\rho =0$, either
$A$ or $B$ has to vanish and $u_p (0)=0$. Thus the singularity problem does
not matter for these states. But for $p$ equal to 0 or -1, $u_0$ or $u_{-1}$
can be finite but non zero at the origin. For them the imposition of the
boundary condition $u(0)=0$ clashes with self adjoint property of the 
Hamiltonian. This problem and the subsequent solution is elaborated below.

Let us factorise (\ref{eq33})
\begin{equation}
({\partial\over{\partial\rho}}-{{p+1}\over \rho})u_p=v_p,~~
({\partial\over{\partial\rho}} +{{p+1}\over \rho})v_p=-u_p, \label{eq36}
\end{equation}
or
\begin{equation}
({\partial\over{\partial\rho}}+{p\over \rho})u_p=v_p,~~
({\partial\over{\partial\rho}}-{p\over \rho})v_p=-u_p, \label{eq37}
\end{equation}
For $p$ taking values 0 or 1, the pairs of equations become identical and
reduce to
\begin{equation}
T\left(\begin{array}{lcr}
v\\u\end{array}\right)=\left(\begin{array}{lcr}
0 & {\partial\over{\partial\rho}}\\ -{\partial\over{\partial\rho}} & 0\end
{array}\right)
\left(\begin{array}{lcr}
v\\u\end{array}\right)=
\left(\begin{array}{lcr}
v\\u\end{array}\right). \label{eq38}
\end{equation}
The deficiency subspace is generated by a pair of
normalised vectors (i.e. $n_+ =n_{-} =1$)
\begin{equation}
T\chi_{\pm}=\pm i\chi_{\pm},~~\chi_{\pm}=e^{-r}
\left(\begin{array}{lcr}
-1\\{\pm i}\end{array}\right). \label{eq39}
\end{equation}
Now the extension of $T$ is $\tau$ where $T=\tau$ with domain
\begin{equation}
D(\tau)=\{\psi + \beta \chi_+ +\gamma\beta\chi_- \mid \psi\epsilon D(T), \beta\epsilon C\}, \label{eq40}
\end{equation}
and the operation of $\tau$ on $D(\tau)$ is
\begin{equation}
\tau (\psi + \beta \chi_+ +\gamma\beta\chi_-)=
T\psi + i\beta \chi_+ -i \gamma\beta\chi_-,
\label{eq41}
\end{equation}
where $\mid {\gamma}\mid $ is an isometry that maps $\chi_+\to\chi_-$. We
parametrise $\gamma =e^{i\alpha}$
and there are diferent extentions for different $\alpha $. In this extended domain
the deficiency indices vanish,
\begin{equation}
n_{\pm}(\tau_\alpha)=0.
\end{equation}
This is the self-adjoint extension explained before.
 The vital role played by the contact term in
the magnetic field in (\ref{eq4}), which gave rise to the ${g\over e}$ term in
$L$ becomes manifest only now. Note that in the expression 
 $p(p+1)=-({1\over 4}-(l-{g\over e})^2)$, since $l$ is an integer, the LHS can
vanish, (i.e., $p=0,-1$), only if the ${g\over e}$ term cancells ${1\over 4}$ in the RHS.

From a slightly different angle, let us now see how the boundary conditions
clash with the self-adjointness of the Hamiltonian at the origin.
Indeed, this is probably an easier way to understand the connection between deficiency
subspaces and the bound state spectrum in question. From the definition of the
self adjoint operator, it follows that the Hamiltonian is self adjoint if
$$(\phi,T\psi )-(T\phi,\psi )=0$$
with $\phi =
\left(\begin{array}{lcr}
u_1\\v_1\end{array}\right)\epsilon D(T^*)$ and
 $\phi =
\left(\begin{array}{lcr}
u_2\\v_2\end{array}\right)\epsilon D(T)$. Using the natural boundary conditions
$u_2(0)=v_2(0)=0$ and $u_2(\infty )=v_2(\infty )=0$, (since these specify $D(T)$),
the above condition reduces to
\begin{equation}
{v_1}^*(0)u_2(0) -{u_1}^*(0)
v_2(0))=0 \label{eq43}
\end{equation}
Thus {\it the above equation
vanishes, independent of any bounday condition on $\phi $, which is in the
domain of $T^*$}. Thus $D(T^*)>D(T)$ since it consists of similar vectors as
in $D(T)$, but without the boundary condition. This makes $T$ not self adjoint
in this domain. The cure is to moderate the too strong boundary condition on $D(T)$
by eg.,
\begin{equation}
u_2(0) +akv_2(0)=u_2(0) +au'_2(0)=0, \label{eq44}
\end{equation}
where $a$ is an arbitrary real parameter. This is the extension $\tau$.  
This ensures {\it identical boundary
conditions on $\psi $ and $\phi $, and hence the domains $D(\tau^*)$ and $D(\tau)$
are identical}. $a$ and $\alpha $ of (\ref{eq41}), (\ref{eq44}),  are related in the following way,
\begin{equation}
a={1\over k}cot{\alpha\over 2}. \label{eq45}
\end{equation}
Now we can get the bound state spectrum easily. For either $p=0$ or $p=-1$
we have
\begin{equation}
 u'=kv, ~~~
v' =-ku, \label{eq46}
\end{equation}
We also have from the extended domain
\begin{equation}
u(0) +au'(0)=0. \label{eq47}
\end{equation}
Using the above equations we get
\begin{equation}
-{1\over k}v'(0) +akv(0) =0. \label{eq48}
\end{equation}
Let us consider some form of $u$ and $v$ which vanish at $r=\infty $. 
\begin{equation}
u=Pe^{-\gamma r}, v=Qe^{-\gamma r}. \label{eq49}
\end{equation}
$P$, $Q$ and $\gamma $ are constants. Solving the above set of equations we find
$$k^2=-{1\over {a^2}}$$,
and so the energy is
\begin{equation}
E=-{1\over{2Ma^2}}. \label{eq50}
\end{equation}

Note that we used the two  component factorised form from (\ref{eq33}) since
the (matrix) differential operator in (\ref{eq38}) is Hermitian in its domain,
although individually $i{d\over {dr}}$ is Hermitian. However, since
${d^2\over{dr^2}}$ is Hermitian, we can use (\ref{eq33}) directly as well.
Choosing $p=0$ or -1 with 

$$u=Ae^{-\gamma r},~~~u(0)+au'(0)=0$$
we again obtain $k^2=-{1\over a^2}$. The difference in these two
formulations is nontrivially manifested in the deficiency index analysis
where the domains of these operators are different.

 Thus (\ref{eq50}) is the cherished expression for the bound state spectrum. These states
correspond to the parameter values such that $L=l-{g\over e}={1\over 2}$ with
integer values for $l$. So far we only know that $a$ is real. There are an 
infinite possibility of different extensions for distinct values of $a$ and
still further conditions are required to uniquely specify $a$. For a zero energy
 bound state, we require $a=\infty $ and
 $u'(0)=0$ {\it but there is no restriction on $u(0)$}.

\section{Conclusion}

We have considered a system of point charge, interacting with a Nielsen-Olesen
vortex of the abelian Higgs model, in 2+1-dimensions. We have shown that classical
bounded orbits are allowed, for certain restricted range of parameters.

Next we quantise the particle motion, treating the potential as external. From
general arguments it is established that in general negative energy bound states
are not present. However, the nature of the effective potential energy allows
metastable states of considerable lifetime.

On the other hand, presence of the contact interaction, (due to the vortex),
makes the analysis subtle for certain angular momenta.
 We have shown that at least for some values of the
parameters,
 at the origin the self adjoint property of the Hamiltonian is lost, which is
restored by the self adjoint extensions, and in this new domain, bound states
appear. There is a one parameter arbitraryness in the bound state energy
spectrum which can be removed by introducing further physical input.

Recently we have come across the papers \cite{os} where the self-adjoint
extensions of the Dirac Hamiltonian have been studied, in the context of
Aharanov-Bohm effect and in the presence of a magnetic vortex.

\vskip1cm
\noindent
{\bf Acknowledgement:}
It is a pleasure to thank Dr. S. Tarafdar and Professor P. Mitra
for many useful discussions. I am also grateful to Professor R.
Jackiw for informing me about the work of C. J. Callias.

\newpage
{\bf Figure Captions}\\
\vskip2cm 
Fig. 1\\
\vskip1cm
$I\to e^x  ~vs. ~x,~II\to \alpha x~ vs.~ x,~III\to \alpha {x\over 4}(x+1)^2
~vs.~ x$, where $x=2eFr,~\alpha ={1\over 2}({g\over {eL}})^2=7.4$
\vskip2cm
Fig. 2\\
\vskip1cm
$V(x)={1\over {x^2}}(1-\sqrt{2\alpha x}e^{-x})^2$ with $\alpha =7.4$  
   \end{document}